\documentclass[conference]{IEEEtran}
\IEEEoverridecommandlockouts
\usepackage[nospace,noadjust]{cite}
\usepackage{amsmath,amssymb,amsfonts}
\usepackage{algorithmic}
\usepackage{graphicx}
\usepackage{subcaption}
\usepackage{textcomp}
\usepackage{xcolor}
\usepackage{tikz}
\usepackage{soul}
\usetikzlibrary{patterns}
\usepackage{pgfplots}
\usetikzlibrary{arrows,shapes,backgrounds,fit}
\usepgfplotslibrary{colorbrewer}
\usetikzlibrary{calc,positioning,shapes,decorations.pathreplacing}
\usepackage[nolist]{acronym}
\usepackage{verbatim}

\usepackage{hyperref}
\hypersetup{colorlinks,allcolors=black}

\usepackage{array}
\usepackage{makecell}

\usepackage{tikz}
\usetikzlibrary{shapes.geometric,arrows}
\tikzstyle{startstop}=[rectangle, minimum width=1.1cm, minimum height=0.1cm,text centered, draw=black]
\tikzstyle{input}=[coordinate]
\tikzstyle{arrow}=[thick,->,>=stealth]

\usepackage{circuitikz}
\usetikzlibrary{positioning}
\usetikzlibrary{shapes,arrows,decorations.pathreplacing}
\tikzset{block/.style = {draw, fill=white, rectangle,
              minimum height=3em, minimum width=1.2cm},
    input/.style = {coordinate},
    output/.style = {coordinate},
    pinstyle/.style = {pin edge={to-,t,black}},
    radiation/.style={decorate,decoration={expanding waves,angle=12,segment length=4pt}}
}

\def\BibTeX{{\rm B\kern-.05em{\sc i\kern-.025em b}\kern-.08em
    T\kern-.1667em\lower.7ex\hbox{E}\kern-.125emX}}
\pgfplotsset{compat=1.16}
\begin{document}

\title{Drone classification from RF fingerprints using deep residual nets\\
\thanks{}
}

\author{\IEEEauthorblockN{Sanjoy Basak\IEEEauthorrefmark{1}, Sreeraj Rajendran\IEEEauthorrefmark{2}, Sofie Pollin\IEEEauthorrefmark{2},  and Bart Scheers\IEEEauthorrefmark{1}}
\\
\IEEEauthorblockA{
Email: \{sanjoy.basak, bart.scheers\}@rma.ac.be,\\
\{sreeraj.rajendran, sofie.pollin\}@esat.kuleuven.be\\
\IEEEauthorrefmark{1}Department CISS, Royal Military Academy, Belgium,
\IEEEauthorrefmark{2}Department ESAT, KU Leuven, Belgium}}
\addtolength{\topmargin}{+0.2cm}
\maketitle

\begin{abstract}
Detecting UAVs is becoming more crucial for various industries such as airports and nuclear power plants for improving surveillance and security measures. Exploiting \ac{rf} based drone control and communication enables a passive way of drone detection for a wide range of environments and even without favourable \ac{los} conditions. In this paper, we evaluate RF based drone classification performance of various \ac{soa} models on a new realistic drone RF dataset. With the help of a newly proposed residual \ac{cnn} model, we show that the drone \ac{rf} frequency signatures can be used for effective classification. The robustness of the classifier is evaluated in a multipath environment considering varying Doppler frequencies that may be introduced from a flying drone. We also show that the model achieves better generalization capabilities under different wireless channel and drone speed scenarios. Furthermore, the newly proposed model's classification performance  is evaluated on a simultaneous multi-drone scenario. The classifier achieves close to 99$\%$ classification accuracy for \ac{snr} 0 dB and at -10 dB \ac{snr} it obtains 5$\%$ better classification accuracy compared to the existing framework.
\end{abstract}


\section{Introduction}
\begin{acronym}[HBCI]
%
%
%
%
%

\acro{3gpp}[3GPP]{3\textsuperscript{rd} Generation Partnership Program}
\acro{cnn}[CNN]{Convolutional Neural Network}
\acro{dnn}[DNN]{Deep Neural Network}
\acro{dl}[DL]{Deep Learning}
\acro{drnn}[DRNN]{Deep Residual Neural Network}
\acro{relu}[RELU]{rectified linear unit}
\acro{selu}[SeLU]{scaled exponential linear units}
\acro{rc}[RC]{radio control}
\acro{fc}[FC]{fully connected}
\acro{rpas}[RPAS]{remotely piloted aircraft systems}
\acro{gap}[GAP]{global  average  pooling}
\acro{gcs}[GCS]{ground control station}
\acro{fbmc}[FBMC]{Filter Bank Multicarrier}
\acro{phy}[PHY]{physical layer}
\acro{pu}[PU]{Primary User}
\acro{rat}[RAT]{Radio Access Technology}
\acro{rfnoc}[RFNoC]{RF Network on Chip}
\acro{sdr}[SDR]{Software Defined Radio}
\acro{su}[SU]{Secondary User}
\acro{toa}[TOA]{Time of Arrival}
\acro{tdoa}[TDOA]{Time Difference of Arrival}
\acro{usrp}[USRP]{Universal Software Radio Peripheral}
\acro{amc}[AMC]{Automatic Modulation Classification}
\acro{lstm}[LSTM]{Long Short Term Memory}
\acro{soa}[SoA]{state-of-the-art}
\acro{fft}[FFT]{Fast Fourier Transform}
\acro{wsn}[WSN]{Wireless Sensor Networks}
\acro{iq}[IQ]{In-phase and quadrature phase}
\acro{snr}[SNR]{signal-to-noise ratio}
\acro{sps}[sps]{samples/symbol}
\acro{awgn}[AWGN]{Additive White Gaussian Noise}
\acro{dsss}[DSSS]{Direct Sequence Spread Spectrum}
\acro{gof}[GoF]{Goodness-of-Fit}
\acro{fhss}[FHSS]{Frequency Hopping Spread Spectrum}
\acro{ofdm}[OFDM]{Orthogonal Frequency Division Multiplexing}
\acro{los}[LOS]{line of sight}
\acro{psd}[PSD]{Power Spectral Density}
\acro{svm}[SVM]{Support Vector Machines}
\acro{xgboost}[XGBoost]{Extreme Gradient Boosting}
\acro{aae}[AAE]{Adversarial Autoencoder}
\acro{vae}[VAE]{Variational Autoencoder}
\acro{saif}[SAIFE]{Spectrum Anomaly Detector with Interpretable FEatures}
\acro{roc}[ROC]{Receiver operating characteristic}
\acro{rnn}[RNN]{Recurrent Neural Network}
\acro{auc}[AUC]{Area Under Curve}
\acro{dsa}[DSA]{Dynamic Spectrum Access}
\acro{hmm}[HMM]{Hidden Markov Models}
\acro{iot}[IoT]{Internet of Things}
\acro{api}[API]{Application Programming Interface}
\acro{gnss}[GNSS]{Global Navigation Satellite System}
\acro{cobras}[COBRAS]{Constraint-based Repeated Aggregation and Splitting}
\acro{ssdo}[SSDO]{Semi-Supervised Detection of Outliers}
\acro{osvm}[OSVM]{One class Support Vector Machine}
\acro{ifo}[IFO]{Isolation Forest}
\acro{lof}[LOF]{Local Outlier Factor}
\acro{rcov}[RCOV]{Robust Covariance}
\acro{rf}[RF]{radio frequency}
\acro{loda}[LODA]{Lightweight on-line detector of anomalies}
\acro{ari}[ARI]{Adjusted Rand Index}
\end{acronym}
Mini \ac{rpas} are imposing threats to national security. In recent years, the threat has become increasingly vivid due to the wide availability of low-cost drones. Several illicit incidents at security-sensitive places such as airports, national campaigns, international sports events and nuclear power plants \cite{drone_threats} have been recorded. The conventional radar, video and acoustic detection systems may fail to detect and identify a drone, due to the small radar cross section, its resemblance to a bird or absence of \ac{los} for vision based schemes, weather or daylight conditions and presence of high noise for acoustic systems \cite{RF_betterthanothers_detectors}. On the contrary, an RF-based detection can work even in NLOS conditions, independent of the object size and daylight or weather condition and is capable to detect a drone from several kilometers away. 

\begin{figure}[ht]
\centering
\begin{minipage}[b]{\linewidth}
\centering
\includegraphics[width=0.6\textwidth]{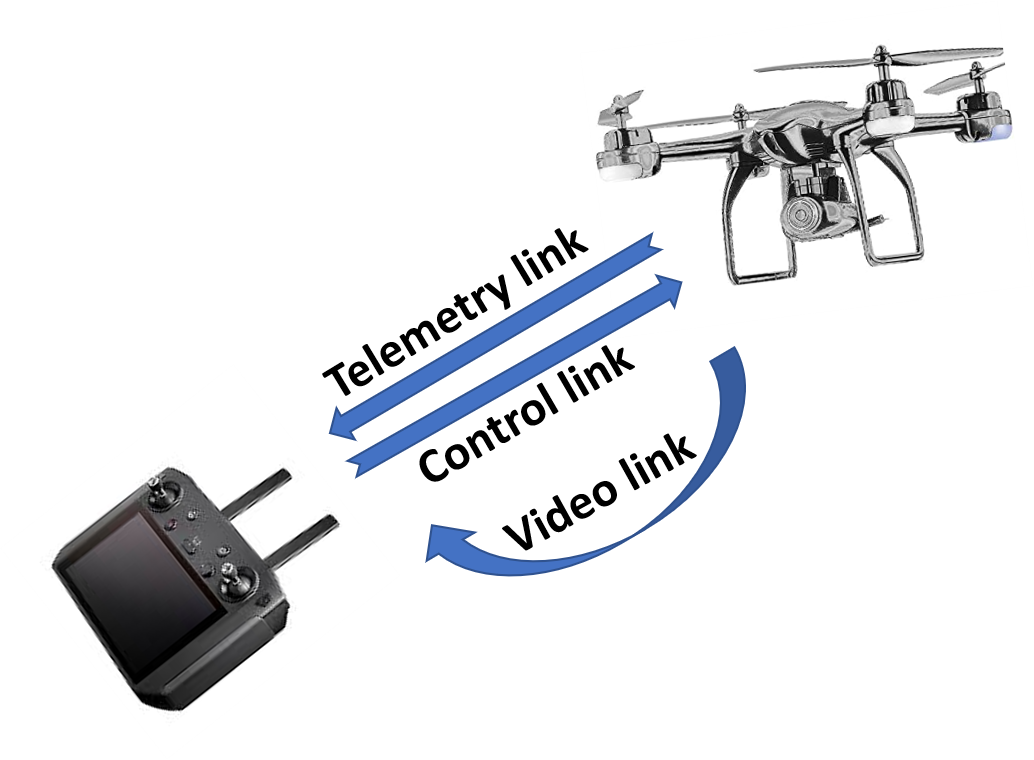}
\end{minipage}
\caption{Drone RF communication}
\label{fig:dronecomm}
\end{figure}

Several RF sensing based drone detection techniques have been proposed in the literature \cite{costeffective_dronedet_Nguyen,dronedet_wifipacketcount_length,dronedatabase_qatarpaper}. A joint active and passive drone detection and tracking system is proposed in \cite{costeffective_dronedet_Nguyen}, where the detection is performed through passive RF sensing at 2.4 GHz. In order to get the \ac{rc} signal and to update its status, most of the drones communicate with the \ac{gcs} at every 33 ms approximately \cite{dronecomm30times1sec_3}, whereas mobile devices and access points (APs) exchange beacons at every 100 ms \cite{wifibeacons}. This criteria was used in \cite{costeffective_dronedet_Nguyen} to detect the presence of a drone. It was found in \cite{costeffective_dronedet_Nguyen} that due to the motor rotation, camera feed, the propeller rotation and the communication channels of a flying drone, an RF emission happens within a frequency range from 20 Hz to 100 Hz. The identification of a drone was performed in \cite{costeffective_dronedet_Nguyen}, by sensing a frequency range of 20 Hz to 100 Hz. In \cite{dronedet_wifipacketcount_length}, the drone identification was performed by measuring the packet length in bytes transmitted by a drone. However, these are inefficient methods, since, (a) any device transmitting at 30 times/sec or transmitting packets of similar length can fool the detector into believing that the signal is coming from a drone, (b) RF signal emission from a drone below 100 Hz is not an active communication signal, therefore, the detection based on this signal will be limited up to a short distance, (c) Multiple drones can transmit signals at the same rate with a similar packet length. Therefore, for robust drone classification, it is important to take into account the time and frequency domain RF signatures of a drone signal. We argue that the physical layer RF signatures are more usable than MAC layer and other passive signal emissions. 

In \cite{dronedatabase_qatarpaper}, the authors utilized frequency domain signatures to detect and classify multiple drones. The authors have used a three layer \ac{fc} \ac{dnn} to detect and identify a drone. The analysis was only performed with three drone signals and the classification of the drone signals was not performed. Alongside the impact of noise and channel variations on the detection and the identification performance was not demonstrated. An efficient classification method is not only necessary for drone localization, this is also required for a smart jammer to neutralize a drone's signal without interrupting other communications. The classification can also provide an insight about the type (e.g fixed wing or quadcopter) and size of drone. The drone communications are performed in the ISM bands and the ISM bands are generally occupied by several heterogeneous sources like WiFi, Bluetooth, ZigBee and several other sources. All transmission sources use spread spectrum modulation techniques like \ac{fhss} and \ac{dsss}. Similarly, the drones also use spread spectrum modulation techniques for the communication and video signal transmission. Since, all sources occupying ISM bands use the same transmission technique, it is difficult to identify and classify a drone's signal. This motivates us to perform an in-depth analysis of the classification of drone signals on a larger dataset containing more commercial drones. We presented the detection of drone RF signals using \ac{gof} based blind spectrum sensing algorithm \cite{GoF_blindsensing}  in \cite{safeshore_passiveradio}. In this paper, we propose a novel RF-based drone classification method.
The main contributions of this paper are the following:
\begin{enumerate}
 \item A novel and realistic dataset is created using nine commercial drone and one WiFi signal. The dataset will be made public for future research\footnote{https://github.com/sanjoy-basak/dronesignals}, once the paper gets accepted.
  \item A new deep residual neural network is proposed to classify single and multi-drone scenarios using frequency RF fingerprints. 
  \item The classification performance under AWGN conditions are evaluated along with insights of the learned features using class activation mapping (CAM) method.
  \item The impact of residual mapping and layer variations on the classification performance are presented. 
  \item The impact of channel variations on the classification performance is analyzed in detail. We show that our model can deal with channel variations and has good generalization ability.
  \item Our model's performance on simultaneous multi-drone classification is also evaluated. We show that our model can classify up to 7 drones simultaneously using the frequency domain RF signatures. As per our knowledge, we are the first to investigate the simultaneous multi-drone classification problem.
\end{enumerate}

The rest of the paper is organized as follows: The challenges associated with RF based drone classification and the received signal from a drone are presented in section \ref{problemstatement}. A mathematical model of the received drone is presented in section \ref{problemstatement}. A brief overview about the wireless signal classification using \ac{dnn} frameworks is presented in section \ref{CNNbasedDNN}. Afterwards, a deep residual neural network is proposed to classify drone based on the time and frequency domain fingerprints. In section \ref{datasetpreparation}, the experimental setup and dataset development procedures are explained. Detailed single and multi drone classification performance analysis are presented in section \ref{resultanalysis}.

\section{Problem statement}\label{problemstatement}
The transmit signal from a drone can be expressed as:
\begin{equation}\label{transmitsig}
\begin{split}
S(t) & = \sqrt{\frac{2E_s}{T_s}} C_k(t) cos(\omega _c t + \phi) \\
& = M(t)cos(\omega _c t + \phi)
\end{split}
\end{equation}

Here, $\frac{E_s}{T_s}$= power of the transmitted symbol,\\
$T_s$ = symbol duration,\\
$C_k(t)$ = spreading code,\\
$M(t)$ = modulated signal,\\
$\omega _{c,\textrm{DSSS}} $ = $2\pi f_c$, \\
$\omega _{c,\textrm{FHSS}} $ = $2\pi (f_l + f_m)$, \\
$\phi$ = phase of the TX signal, \\
$f_l$ = intermediate freq for the selected band of the $l^{\mathrm{th}}$ hop, \\
$f_m$ = $m^{\mathrm{th}}$ significant freq. \\

Generally, $C_k(t)$ is only used in the DSSS signal. The DSSS transmitters generally scans the radio spectrum to find an unoccupied channel and perform communication with the selected band. On the contrary, a FHSS transmitter transmits with a frequency ranging between $f_l-f_m$ and $f_l+f_m$.

The received signal in a multipath fading scenario can be expressed as
\begin{equation}\label{recvsig}
y(t) = AM(t)cos(\omega _c t) + \sum_{n=2}^{n=N} M(t) r_n cos(\omega _ct + \phi _n)
\end{equation}

Here, A = amplitude of the LOS component,\\
$r_n$ = amplitude of the n$^{\mathrm{th}}$ reflected wave,\\
$\phi _n$ = phase of the n$^{\mathrm{th}}$ reflected wave,\\
n = 2, 3, ..., N denotes the reflected and scattered waves.\\

The time domain RX signal is broken into $M_t$ consecutive segments, each with length $N_f$ (= FFT size). We perform DFT on each segment and convert the signal into frequency domain. This gives us a DFT matrix $Y(\tau,\omega )$ of size $M_t \times N_f$. The magnitude of $Y(\tau,\omega )$ gives the spectrogram matrix. The aim is to classify the drone signal from the spectrogram matrix.

\section{DNN based wireless signal classification}\label{CNNbasedDNN}
\begin{figure}[ht]
\centering
\begin{minipage}[b]{0.40\linewidth}
\centering
\begin{tikzpicture}[node distance =1.3cm]

\node [input, name=input] {};
\node(start)[startstop,below of=input,yshift=0.5cm,align=center]{\small 3x3 conv,\\ \small RELU};
\node(convlinear)[startstop,below of=start,yshift=0.2cm,align=center]{\small 3x3 conv};
\node(pluscircle)[draw,fill=white,circle,below of=convlinear,yshift=0.2cm]{+};
\node (activation)[startstop, below of=pluscircle,yshift=0.3cm,align=center] {\small RELU};

\draw [arrow] (input) -- (start);
\draw[arrow](start)--(convlinear);
\draw[arrow](convlinear)--(pluscircle);
\draw[thick,->,>=stealth] (pluscircle)--(activation);
\draw[arrow] (input) to [bend left=90] node[anchor=south west,align=center] {\small skip\\ \small connection} (pluscircle);
\end{tikzpicture} 
\subcaption{ResUnit}
\label{fig:resunit}
\end{minipage}
\hfill
\begin{minipage}[b]{0.45\linewidth}
\centering
\begin{tikzpicture}[node distance =1.5cm]
\node [input, name=input] {};
\node(start)[startstop,below of=input,yshift=0.6cm,align=center]{\small 1x1 Conv, \\ \small Linear};
\node(resunit1)[startstop,below of=start,yshift=0.3cm]{\small ResUnit};
\node(resunit2)[startstop,below of=resunit1,yshift=0.4cm]{\small ResUnit};
\draw [arrow] (input) -- (start);
\draw [arrow] (start) -- (resunit1);
\draw [arrow] (resunit1) -- (resunit2);
\end{tikzpicture} 
\subcaption{Residual stack}
\label{fig:residualstack}
\end{minipage}
\caption{Building blocks of the deep residual network architecture}
\label{fig:deep_resnet_architecure}
\end{figure}
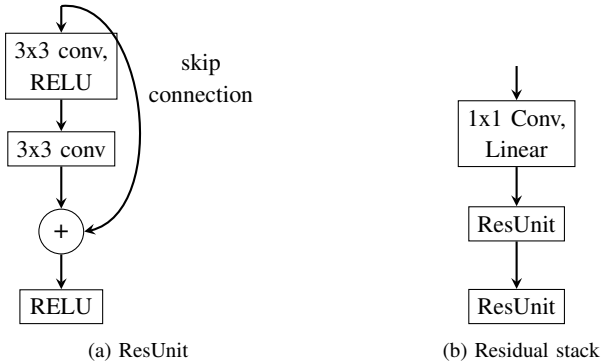

\begin{table}[htbp]
\centering
\caption{DRNN architecture}
\begin{center}
\begin{tabular}{|c|c|}
\hline
{Layer} & {Output dimensions}\\
\hline
Input & 256 x 256\\
\hline
Residual stack & 256 x 256 x 32\\
\hline
MP & 128 x 128 x 32\\
\hline
Residual stack & 128 x 128 x 32\\
\hline
MP & 64 x 64 x 32\\
\hline
Residual stack & 64 x 64 x 32\\
\hline
MP  & 32 x 32 x 32\\
\hline
Residual stack & 32 x 32 x 32\\
\hline
MP  & 16 x 16 x 32\\
\hline
Residual stack & 16 x 16 x 32\\
\hline
GAP & 32\\
\hline
FCL/Softmax & 10 \\
\hline
\end{tabular}
\label{tab:yoloarch}
\end{center}
\end{table}

The potential of wireless signal classification has been widely investigated in the automatic modulation classification problem in the last few years \cite{lstm_classif,cnnresnet_oshea,oshea_cnn,cnnparallelerma}. These studies showed that there are several advantages of using a \ac{dl} based classifier over a conventional feature extractor. A DL classifier can identify a radio signal from the time and frequency signatures of raw input signals without extracting any signal specific expert features \cite{lstm_classif,cnnresnet_oshea} from the signal. The classification can be performed in a supervised or unsupervised manner. A supervised learning method learns from a perfectly labelled dataset and later, it can be deployed for classification based on label prediction. On the contrary, an unsupervised learning method does not require any labelled dataset \cite{unsupervised_fingerprinting}.

\subsection{CNN based deep neural networks}
The state of art \ac{cnn} architectures have shown superior performance in image and speech recognition problems \cite{cnn_image_detection,cnn_speech_recognition}. For wireless signal recognition and device fingerprinting, the \ac{cnn} architectures showed great performances too \cite{radio_fingerprinting,oshea_cnn,cnnresnet_oshea}. In this section, a brief introduction of the core layers of the CNN architectures is provided. 

Convolution is the most vital operation in a \ac{cnn} architecture. A convolution layer consists of a set of filters and it performs the convolution operation to extract features from the input dataset. A convolution is generally followed by an activation operation such as \ac{relu}. If the input is defined by x and the output by y, then the \ac{relu} activation can be shown as:
\begin{equation}
y=\textrm{max}(0, x \times W+b)\label{relueq}
\end{equation}
Here, W denotes the learned weight matrix and b denotes the bias vector.

\subsection{Deep residual neural network}
A deep CNN architecture can extract more complex features by increasing the number of stacked layers compared to its shallower counterpart. However, this creates the problem of vanishing or exploding gradients \cite{ref1_resnet_gradientdegradation,ref2_resnet_gradientdegradation}. The vanishing/exploding gradient problem has been solved by the normalization of the input data (i.e. normalized initialization) \cite{ref2_resnet_gradientdegradation} and the intermediate layers (i.e. batch normalization) \cite{batch_normalization} of a \ac{dnn}. However, even after these normalizations, a degradation problem was observed in \cite{resnetpaper}, with the increase of network depth. This results in the saturation and rapid degradation of classification accuracy. One prominent approach to solve this degradation problem is through the use of a residual network \cite{resnetpaper}. The building block of a residual network is an identity mapping operation with its residual path. The network performs an addition of the output of a two-layer network with its input. The operation can be shown as:
\begin{equation}
y= F(x,{W_i}) + x. \label{residual1eq}
\end{equation}

Here, x is the input and $F(x,{W_i})$ is the residual mapping of the network. 

For the classification of a drone using RF fingerprints, we propose a \ac{drnn}, which is an adaptation of the residual neural network proposed in \cite{cnnresnet_oshea}. To visualize the class activation, a \ac{gap} layer as proposed in \cite{activation_map_GAP} is utilized. The residual unit and the residual stack are shown in Fig \ref{fig:resunit} and \ref{fig:residualstack} respectively. Two 2D convolution operations with kernel size 3x3 are performed in the residual unit. Alongside, RELU activation is used after the first convolution and after the skip connection as shown in Fig \ref{fig:resunit}. The residual stack consists of a convolution operation with a kernel size of 1x1 and linear activation followed by two residual units. 

The complete \ac{drnn} architecture is shown in Table \ref{tab:yoloarch}. For each residual stack, we have used a filter size of 32. The network architecture consists of 5 residual stacks. Each residual stack is followed by a max-pooling (MP) layer with kernel size 2x2 apart from the final residual stack. The final residual stack is followed by a GAP layer. Finally, a fully connected layer (FCL) is used with a softmax activation to get the prediction probability. For simultaneous multi-drone signal classification, we have used sigmoid activation instead of softmax activation.

The class activation map is produced using a GAP in our \ac{drnn} model as defined in \cite{activation_map_GAP}. Let $f_k(x,y)$ represents the activation of unit k, in the last convolution layer (before the GAP layer) at a spatial location (x,y). If $w_k^c$ represents the weight for a particular class c for unit k of the last FCL, then the activation map $M_c$ of class c can be defined as:

\begin{equation} \label{gradcam}
M_c (x,y)= \sum_{k} w_k^c f_k (x,y) 
\end{equation}

\section{Drone RF database development}\label{datasetpreparation}
The lack of a large open-source drone database hinders the development of robust RF drone classification techniques. At the time of writing this paper, only one open database \cite{dronedatabase_qatarpaper} exists in literature, containing only three drone signals. For this study, we have created a larger database containing nine drone signals and WiFi signals. The drone signals include RC and video signals at 2.4 GHz, as given in Table \ref{tab:listdronescontr}.

\subsection{Experimental setup}
\begin{figure}[htbp]
\centering
\begin{tikzpicture}[auto,minimum width=1.0cm, minimum height=0.1cm, node distance=1cm,>=latex']
\node[block,align=center](tx){\small Drone \\ \small GCS};
\node[antenna] at (tx.east) {};

\node[block,right = 3.5cm of tx](rx){\small USRP};
\node[antenna,xscale=-1] at (rx.west) {};

\draw[radiation] ([shift={(1cm,2cm)}]tx.east)-- node [above=5mm] {} ([shift={(-1cm,2cm)}]rx.west);

\node[block,right = 0.4cm of rx,align=center](recording){\small Spectrum\\ \small sensing};

\node[block,below = 0.3cm of recording,align=center](freqcenter){\small Introduce\\\small AWGN + \\\small Multipath};

\node[startstop,left = 0.3cm of freqcenter,align=center](downsamp){\small Spectrogram\\ \small dataset\\ \small 256x256};
\node[startstop,left = 0.3cm of downsamp](multi){\small DNN};
\draw [arrow] (rx) -- (recording);
\draw [arrow] (recording) -- (freqcenter);
\draw [arrow] (freqcenter) -- (downsamp);
\draw [arrow] (downsamp) -- (multi);

\end{tikzpicture}
\caption{Measurement setup schematic. The recording was performed in the anechoic chamber. Later AWGN and multipath fading were introduced in the simulation environment.}
\label{fig:experimentalsetup}
\end{figure}
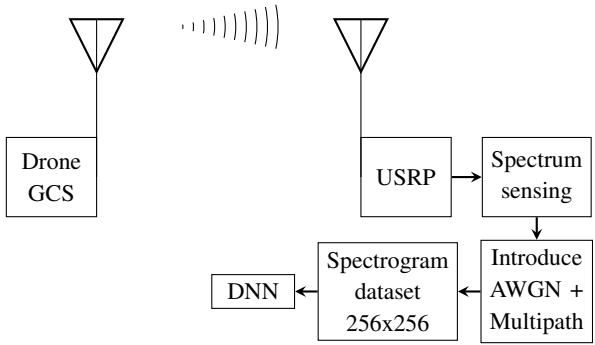

The drone signal was recorded in an anechoic chamber with a universal software radio peripheral (USRP) X310 with a sampling rate of 100 MSps. A computer was connected to the USRP through a 10 Gbit Ethernet cable. An omnidirectional antenna, OmniLOG 70600, was used with the USRP to receive signal. Nine commercial radio controllers and drones and one WiFi router were used for the data collection process as given in Table \ref{tab:listdronescontr}. The drones and controllers were placed 7 meters apart from the antenna in the anechoic chamber and the recordings were made in presence of active communication between them.

\begin{table}[htbp]
\centering
\caption{Drones and Radio Controllers}
\begin{center}
\begin{tabular}{|c|c|c|c|}
\hline
Nr. & Name & {Signal Type} & Freq(GHz)\\
\hline
1 & Parrot Disco & RC+Video & 2.4\\
\hline
2 & Q205 & RC & 2.4\\
\hline
3 & Tello & RC+Video & 2.4\\
\hline
4 & MultiTx & RC & 2.4\\
\hline
5 & Nine Eagles & RC & 2.4\\
\hline
6 & Spektrum DX4e & RC & 2.4\\
\hline
7 & Spektrum DX6i & RC & 2.4\\
\hline
8 & Wltoys & RC & 2.4\\
\hline
9 & S500  & RC & 2.4\\
\hline
10 & Linksys router  & IEEE802.11b/g & 2.4\\
\hline
\end{tabular}
\label{tab:listdronescontr}
\end{center}
\end{table}

\subsection{Training data preparation}
The procedure for the training data preparation is shown in Fig \ref{fig:experimentalsetup}. The signal from the USRP was received in burst mode. After receiving the signal, we performed spectrum sensing to decide if the received burst contains any signal or not. The receive burst was stored if a signal was present, or else it was discarded. Later, in the simulation environment we have introduced AWGN to the signal. The signal coming from the AWGN channel was converted to the frequency domain to prepare the spectrogram dataset of size 256$\times$256. This spectrogram data is used as the training and test dataset in this paper.  

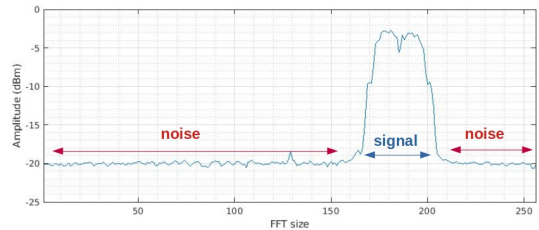
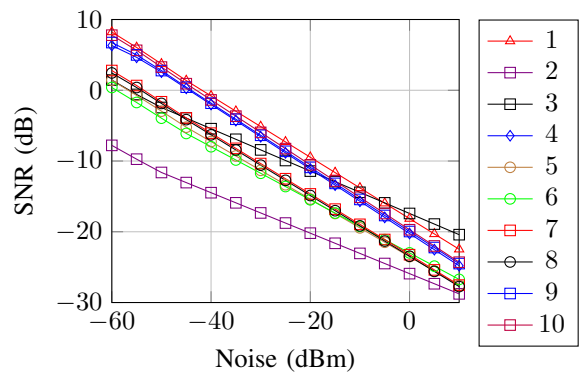
\begin{figure}[htbp]
\centering
\begin{minipage}[b]{0.8\linewidth}
\centering
\includegraphics[trim=2cm 3cm 3.7cm 3.3cm,clip=true,width=\columnwidth]{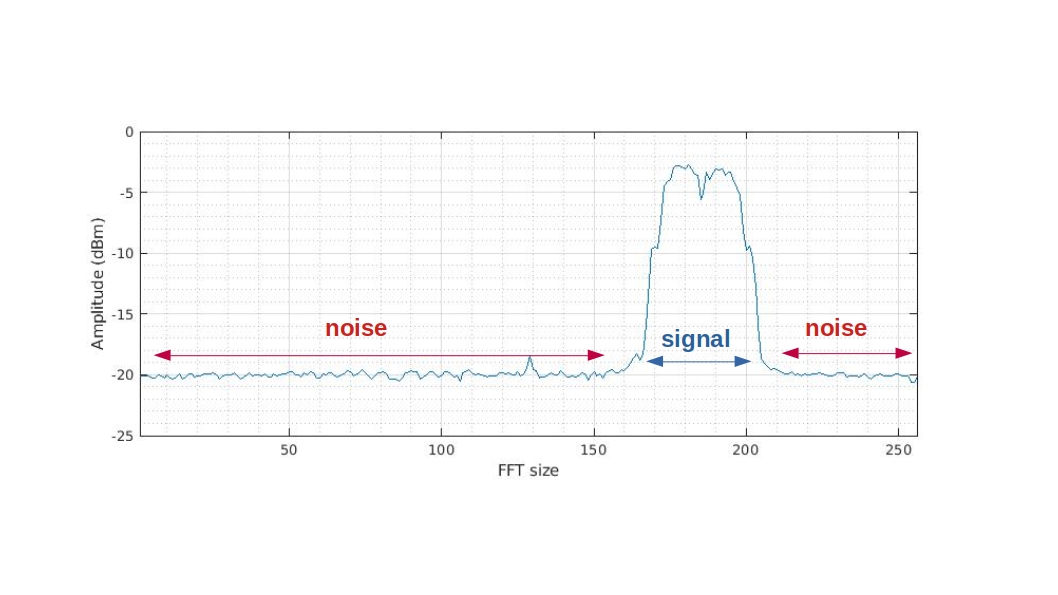}
\subcaption{SNR calculation in frequency domain}
\label{fig:snrenergy}
\end{minipage}
\par\bigskip
\begin{minipage}[b]{0.7\linewidth}
\centering
\begin{tikzpicture}
\begin{axis}
[legend style={at={(1.2,1)},anchor=north},legend columns=1,
 width=\columnwidth,grid=both,ylabel=SNR (dB),xlabel=Noise (dBm),grid style={line width=.1pt, draw=gray!10},major grid style={line width=.2pt,draw=gray!50},xmin=-60,xmax=10,ymax=10, ymin=-30,]

\addplot [mark=triangle, color=red] table [x=snr, y=s5, col sep=comma] {results/tikzplots/results_new/snr_vs_noise.csv};\addlegendentry{$1$}

\addplot [mark=square, color=violet] table [x=snr, y=s6, col sep=comma] {results/tikzplots/results_new/snr_vs_noise.csv};\addlegendentry{$2$}

\addplot [mark=square, color=black] table [x=snr, y=s8, col sep=comma] {results/tikzplots/results_new/snr_vs_noise.csv};\addlegendentry{$3$}

\addplot [mark=diamond, color=blue] table [x=snr, y=s3, col sep=comma] {results/tikzplots/results_new/snr_vs_noise.csv};\addlegendentry{$4$}

\addplot [mark=o, color=brown] table [x=snr, y=s4, col sep=comma] {results/tikzplots/results_new/snr_vs_noise.csv};\addlegendentry{$5$}

\addplot [mark=o, color=green] table [x=snr, y=s1, col sep=comma] {results/tikzplots/results_new/snr_vs_noise.csv};\addlegendentry{$6$}

\addplot [mark=square, color=red] table [x=snr, y=s2, col sep=comma] {results/tikzplots/results_new/snr_vs_noise.csv};\addlegendentry{$7$}

\addplot [mark=o, color=black] table [x=snr, y=s10, col sep=comma] {results/tikzplots/results_new/snr_vs_noise.csv};\addlegendentry{$8$}

\addplot [mark=square, color=blue] table [x=snr, y=s7, col sep=comma] {results/tikzplots/results_new/snr_vs_noise.csv};\addlegendentry{$9$}

\addplot [mark=square, color=purple] table [x=snr, y=s9, col sep=comma] {results/tikzplots/results_new/snr_vs_noise.csv};\addlegendentry{$10$}

\end{axis}
\end{tikzpicture} 
\subcaption{Relation between noise vs SNR}
\label{fig:noisevsnrs}
\end{minipage}
\caption{SNR calculations of wideband signals}
\end{figure}

Generally, all sources at 2.4 GHz transmits with a power of approximately 20 dBm, however, the received SNR are different for different signals. This is due to the fact that we kept the RX bandwidth fixed (i.e. 100 MHz) and the bandwidth of the transmission signals are different for different signals. To vary the signal SNR, AWGN ranging from -60 dBm to +10 dBm was added in matlab. The SNR is calculated from the ratio of signal and noise energy in frequency domain as shown in Fig \ref{fig:snrenergy}. The relation between the AWGN and SNR for different signals is shown in Fig \ref{fig:noisevsnrs}. As it can be seen from the figure, for a constant noise level the SNR is different for different signals.

\subsection{Implementation details}
The classifier was implemented with Keras running on top of Tensorflow \cite{keras_ref}. We performed the classification on a Intel core i7 computer with nvidia geforce RTX 2080 GPU. For the optimization, we used adam optimizer \cite{adamoptimizer} with a learning rate of 0.001. In order to calculate the loss function while training, we used categorical crossentropy as the metric. The training was performed for 200 epochs with a batch size of 32.

\section{Results}\label{resultanalysis}

\subsection{Classification under AWGN conditions}
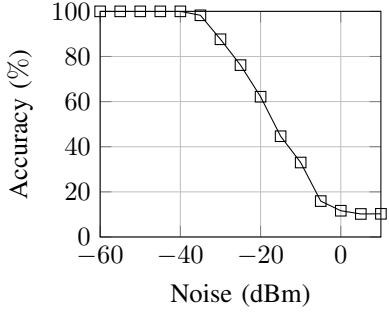
\begin{figure}[t]
\centering
\begin{minipage}[b]{0.6\linewidth}
\centering
\begin{tikzpicture}
\begin{axis}
[legend style={at={(0.45,1.4)},anchor=north},legend columns=3,
 width=\columnwidth,grid=both,ylabel=Accuracy ($\%$),xlabel=Noise (dBm),grid style={line width=.1pt, draw=gray!10},major grid style={line width=.2pt,draw=gray!50},xmin=-60,xmax=10,ymax=100, ymin=0,]

\addplot [mark=square, color=black] table [x=snr, y=acc, col sep=comma] {results/tikzplots/results_new/clc_un_awgn_R4.csv};
\end{axis}
\end{tikzpicture} 
\end{minipage}
\caption{Classification accuracy under AWGN conditions. Train samples: 15k, Test samples: 4k}
\label{fig:clc_awgn_acc}
\end{figure}

\begin{figure}[t]
\centering
\begin{minipage}[b]{0.8\linewidth}
\includegraphics[trim=0.12cm 0.12cm 0.12cm 0.12cm,clip=true,width=\columnwidth]{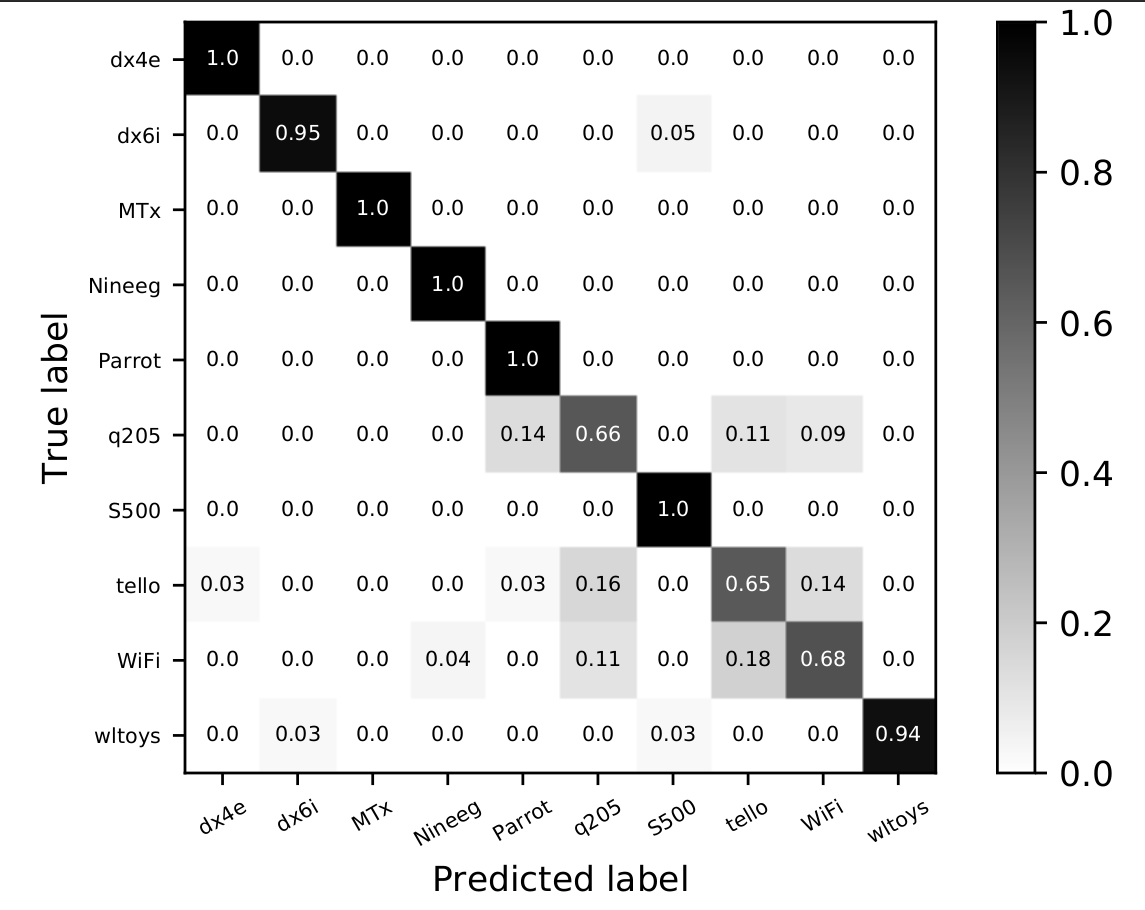}
\end{minipage}
\caption{Confusion matrix with AWGN -30 dBm}
\label{fig:confmatrix}
\end{figure}

The classification accuracy under the AWGN conditions is plotted in Fig \ref{fig:clc_awgn_acc}. We obtained a classification accuracy of nearly 100$\%$ with the introduced noise ranging from -60 to -40 dBm, which corresponds to the average signal SNR of 3 dB to -5 dB (Fig \ref{fig:noisevsnrs}). The accuracy linearly decreases as the noise increases from -40 to 0 dBm. To visualize the classification accuracy of each class, we plotted the confusion matrix at -30 dB noise in Fig \ref{fig:confmatrix}, where we obtained the classification accuracy of 87.6$\%$. At this SNR, we can observe some confusion in the classification between Tello and WiFi, since both use similar transmit signal. Apart from these two TXs, Q205 also shows lower classification accuracy and displays some confusion with Tello and WiFi. Although q205 does not transmit the same type of signal as Tello and WiFi, the transmit power was observed to be lower compared to the other transmitters. Therefore, with higher noise value the signal from q205 goes much inside noise compared to the other devices. 
Overall, the classifier provided good performance, it classified the drone signals even at lower SNR regions and distinguished the drone signals from the WiFi signal. 

\subsection{Effect of multipath environment}
\begin{figure}[htbp]
\centering
\begin{minipage}[b]{0.80\linewidth}
\centering
\begin{tikzpicture}
\begin{axis}
[legend style={at={(0.95,1)},anchor=north, font=\small},legend columns=1,
 width=\columnwidth,grid=both,ylabel=Accuracy ($\%$),xlabel=Noise (dBm),grid style={line width=.1pt, draw=gray!10},major grid style={line width=.2pt,draw=gray!50},xmin=-60,xmax=10,ymax=100, ymin=0,]

\addplot [mark=o, color=green] table [x=snr, y=s1, col sep=comma] {results/tikzplots/results_new/clc_rician_doppler.csv};\addlegendentry{$100 Hz$}

\addplot [mark=square, color=red] table [x=snr, y=s2, col sep=comma] {results/tikzplots/results_new/clc_rician_doppler.csv};\addlegendentry{$500 Hz$}

\addplot [mark=diamond, color=blue] table [x=snr, y=s3, col sep=comma] {results/tikzplots/results_new/clc_rician_doppler.csv};\addlegendentry{$1 kHz$}

\addplot [mark=o, color=brown] table [x=snr, y=s4, col sep=comma] {results/tikzplots/results_new/clc_rician_doppler.csv};\addlegendentry{$50 kHz$}

\addplot [mark=triangle, color=red] table [x=snr, y=s5, col sep=comma] {results/tikzplots/results_new/clc_rician_doppler.csv};\addlegendentry{$100 kHz$}

\addplot [mark=square, color=violet] table [x=snr, y=s6, col sep=comma] {results/tikzplots/results_new/clc_rician_doppler.csv};\addlegendentry{$500 kHz$}

\addplot [mark=square, color=black] table [x=snr, y=s7, col sep=comma] {results/tikzplots/results_new/clc_rician_doppler.csv};\addlegendentry{$1 MHz$}

\end{axis}
\end{tikzpicture}
\subcaption{Classification performance in Rician environment}
\label{fig:multi_ric}
\end{minipage}
\par\bigskip
\begin{minipage}[b]{0.80\linewidth}
\centering
\begin{tikzpicture}
\begin{axis}
[legend style={at={(0.98,1)},anchor=north, font=\small},legend columns=1,
 width=\columnwidth,grid=both,ylabel=Accuracy ($\%$),xlabel=Noise (dBm),grid style={line width=.1pt, draw=gray!10},major grid style={line width=.2pt,draw=gray!50},xmin=-60,xmax=10,ymax=100, ymin=0,]

\addplot [mark=o, color=green] table [x=snr, y=s1, col sep=comma] {results/tikzplots/results_new/clc_rayleigh_doppler.csv};\addlegendentry{$100 Hz$}

\addplot [mark=square, color=red] table [x=snr, y=s2, col sep=comma] {results/tikzplots/results_new/clc_rayleigh_doppler.csv};\addlegendentry{$500 Hz$}

\addplot [mark=diamond, color=blue] table [x=snr, y=s3, col sep=comma] {results/tikzplots/results_new/clc_rayleigh_doppler.csv};\addlegendentry{$1 kHz$}

\addplot [mark=o, color=brown] table [x=snr, y=s4, col sep=comma] {results/tikzplots/results_new/clc_rayleigh_doppler.csv};\addlegendentry{$10 kHz$}

\addplot [mark=triangle, color=red] table [x=snr, y=s5, col sep=comma] {results/tikzplots/results_new/clc_rayleigh_doppler.csv};\addlegendentry{$20 kHz$}

\addplot [mark=square, color=violet] table [x=snr, y=s6, col sep=comma] {results/tikzplots/results_new/clc_rayleigh_doppler.csv};\addlegendentry{$50 kHz$}

\addplot [mark=square, color=blue] table [x=snr, y=s7, col sep=comma] {results/tikzplots/results_new/clc_rayleigh_doppler.csv};\addlegendentry{$100 kHz$}

\addplot [mark=square, color=black] table [x=snr, y=s8, col sep=comma] {results/tikzplots/results_new/clc_rayleigh_doppler.csv};\addlegendentry{$500 kHz$}

\end{axis}
\end{tikzpicture}
\subcaption{Classification performance in Rayleigh environment}
\label{fig:multi_ray}
\end{minipage}
\caption{Influence of multipath environment on classification performance. The classifier is trained with AWGN faded dataset and tested with Rician and Rayleigh faded dataset.}
\label{fig:multipathprop}
\end{figure}
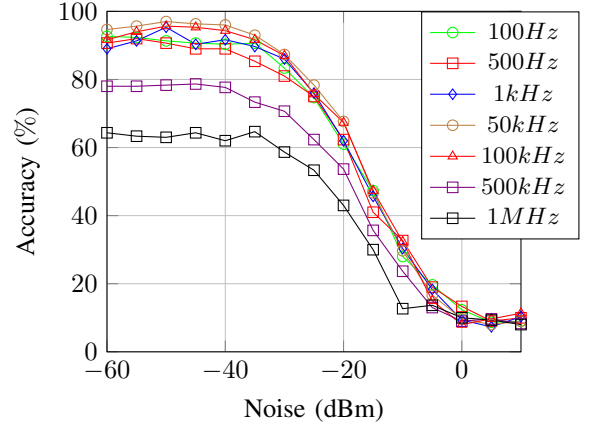
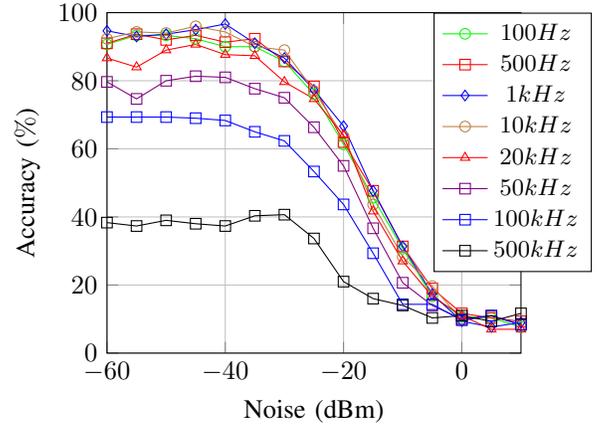


\begin{figure*}[htbp]
\centering
\begin{minipage}[b]{0.19\textwidth}
\centering
\includegraphics[trim=0.5cm 0.5cm 0.5cm 0.2cm,clip=true,width=0.95\textwidth]{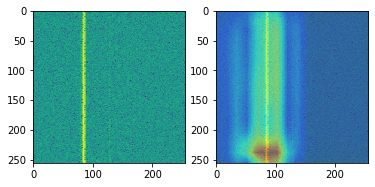}
\subcaption{DX4E}
\label{fig:dx4e_true}
\end{minipage}
\hfill
\begin{minipage}[b]{0.19\linewidth}
\centering
\includegraphics[trim=0.5cm 0.5cm 0.5cm 0.5cm,clip=true,width=0.95\textwidth]{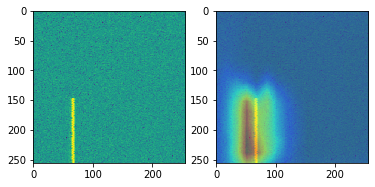}
\subcaption{DX6i}
\label{fig:dx6i_true}
\end{minipage}
\hfill
\begin{minipage}[b]{0.19\linewidth}
\centering
\includegraphics[trim=0.5cm 0.5cm 0.5cm 0.2cm,clip=true,width=0.95\textwidth]{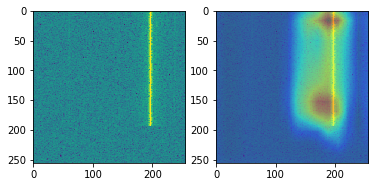}
\subcaption{MultiTx}
\label{fig:multitx_true}
\end{minipage}
\hfill
\begin{minipage}[b]{0.19\linewidth}
\centering
\includegraphics[trim=0.5cm 0.5cm 0.5cm 0.2cm,clip=true,width=\textwidth]{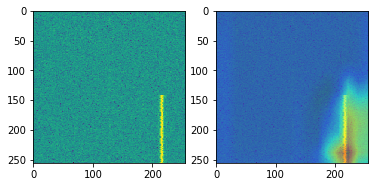}
\subcaption{Nine Eagles}
\label{fig:nineeg_true}
\end{minipage}
\hfill
\begin{minipage}[b]{0.19\linewidth}
\centering
\includegraphics[trim=0.5cm 0.5cm 0.5cm 0.6cm,width=0.95\textwidth]{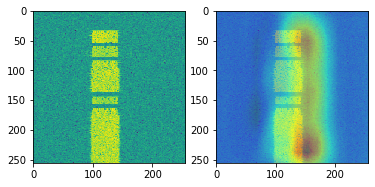}
\subcaption{Parrot}
\label{fig:Parrot_true}
\end{minipage}
\par\bigskip
\begin{minipage}[b]{0.19\linewidth}
\centering
\includegraphics[trim=0.5cm 0.5cm 0.5cm 0.5cm,width=0.95\textwidth]{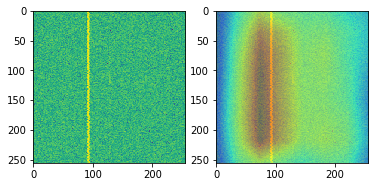}
\subcaption{Q205}
\label{fig:q205_true}
\end{minipage}
\hfill
\begin{minipage}[b]{0.19\linewidth}
\centering
\includegraphics[trim=0.5cm 0.5cm 0.5cm 0.5cm,width=0.95\textwidth]{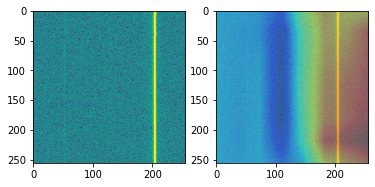}
\subcaption{S500 True}
\label{fig:S500_true}
\end{minipage}
\hfill
\begin{minipage}[b]{0.19\linewidth}
\centering
\includegraphics[trim=0.5cm 0.5cm 0.5cm 0.5cm,width=0.95\textwidth]{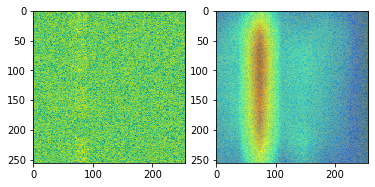}
\subcaption{Tello}
\label{fig:tello_true}
\end{minipage}
\hfill
\begin{minipage}[b]{0.19\linewidth}
\centering
\includegraphics[trim=0.5cm 0.5cm 0.5cm 0.5cm,width=0.95\textwidth]{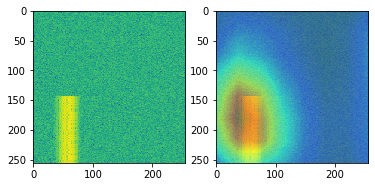}
\subcaption{WiFi}
\label{fig:WiFi_true}
\end{minipage}
\hfill
\begin{minipage}[b]{0.19\linewidth}
\centering
\includegraphics[trim=0.5cm 0.5cm 0.5cm 0.5cm,width=0.95\textwidth]{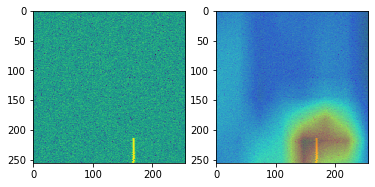}
\subcaption{wltoys}
\label{fig:wltoys_true}
\end{minipage}
\caption{Class activation map for all signals under different AWGN conditions.}
\label{fig:gradcams}
\end{figure*}

To understand the influence of channel variations on the classification performance, we tested the classifier with Multipath faded dataset, which was trained with AWGN conditions. In order to introduce multipath fading, we propagated the signals through a simulated Rician and Rayleigh channel. The following path delay and gain vector was used during the simulation:\newline
path delay (sec) : [0, 200, 800]*$10^{-9}$ \newline
gain vector (dB) : [0, -3, -5] \newline
K-factor: 2.8 \newline
We used Jakes Doppler spectrum for both Rician and Rayleigh channels. For the Rician channel, we used K-factor of 2.8 and the doppler for LOS path was kept the same as for multipath. Later, the signal SNR was varied and the performance was tested.

The classification performance in the Rician multipath channel is shown in Fig \ref{fig:multi_ric}. We can see that the classification performance remained almost the same for the Doppler frequency ranging from 100 Hz to 100 kHz. Although at the higher SNR regions, a slight degradation of around 5-10 $\%$ in the classification accuracy can be observed, the classification performance remained the same for lower the SNR region. Significant drops in the classification accuracy were observed for very high Doppler frequencies like 500 kHz and 1 MHz. At 2.4 GHz, a Doppler frequency of 500 kHz corresponds to a velocity of 6.2457$\times$10$^4$ m/s, which is practically not possible from a commercial drone.

The classification performance in the Rayleigh multipath channel is shown in Fig \ref{fig:multi_ray}. The classifier also showed a good performance with the Rayleigh faded dataset for Doppler frequency ranging from 100 Hz to 10 kHz. The classification performance starts degrading from the Doppler frequency of 50 kHz. We can also clearly see that the classification performance is better with Rician faded dataset compared to the Rayleigh counterpart for higher Doppler frequencies. This is generally expected since for Rayleigh fading the multipath signals does not include the LOS component, the resultant signal suffers more fading. 

Overall, the classifier performed equally well for both Rician and Rayleigh fading scenario for Doppler frequencies corresponding to a real drone's velocity. The performance clearly explains the generalization capability of our classifier.

\subsection{Class Activation mapping}
The activation maps of the drone and WiFi signals are shown in Fig \ref{fig:gradcams}. It provides an insight into what features the model is looking for while activating a particular class. The activation mapping can also be used as a tool for frequency localization in this case. It can be observed from the heatmap that the activation location corresponds to the frequency location of the signal. The model only learns the part of the spectrum which are important for classification. For example, for Nine Eagles (Fig \ref{fig:nineeg_true}), the model learns bottom and edges of the spectrum, whereas for MultiTx (Fig \ref{fig:multitx_true}), the model learns the start and ending of the spectrum.

\subsection{Impact of residual mapping}
\begin{figure}[t]
\centering
\begin{tikzpicture}
\begin{scope}[xshift=0.0cm,yshift=0.0cm]
\begin{axis}[legend style={at={(1.2,1.6)},anchor=north},legend columns=2,width=0.5\columnwidth,grid=both,ylabel=Accuracy (\%) ,xlabel=Noise (dBm),grid style={line width=.1pt, draw=gray!10},major grid style={line width=.2pt,draw=gray!50},xmin=-60,xmax=10,ymax=100, ymin=0,minor tick num=5,legend cell align={left},colormap/hot,,]
\addplot[color= blue,mark size=2pt, mark=square] table [x=snr, y=s1, col sep=comma] {results/tikzplots/results_new/residual_vs_no.csv};
\addlegendentry{DRNN}

\addplot[color=violet,mark size=3pt, mark=diamond] table [x=snr, y=s2, col sep=comma] {results/tikzplots/results_new/residual_vs_no.csv};
\addlegendentry{W/O skip connection}

\end{axis}
\end{scope}
\begin{scope}[xshift=4.5cm,yshift=0.0cm]
\begin{axis}[legend style={at={(0.5,1.2)},anchor=north},legend columns=2,width=0.5\columnwidth,grid=both,ylabel=Accuracy (\%),xlabel=Noise (dBm),grid style={line width=.1pt, draw=gray!10},major grid style={line width=.2pt,draw=gray!50},xmin=-60,xmax=-15,ymax=100, ymin=0,minor tick num=5,legend cell align={left},colormap/hot,,]
\addplot[color= blue,mark size=2pt, mark=square] table [x=snr, y=s1, col sep=comma] {results/tikzplots/results_new/residual_vs_no_lessnoise.csv};
\addlegendentry{DRN}
\addplot[color=violet,mark size=3pt, mark=diamond] table [x=snr, y=s2, col sep=comma] {results/tikzplots/results_new/residual_vs_no_lessnoise.csv};
\addlegendentry{W/O skip connection}

\legend{};
\end{axis}
\end{scope}
\end{tikzpicture}
\caption{Classification performance with and without residual mapping}
\label{fig:residual_perfcompBoth}
\end{figure}
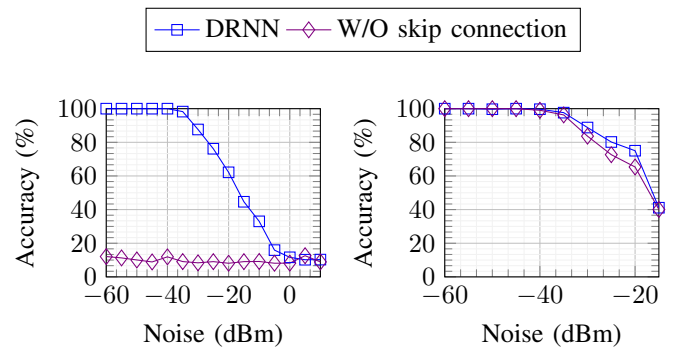

The importance of residual mapping on the classification performance is investigated in this section. To perform the test, we have used 6 layered residual network with and without skip connection. The classification performance for two scenarios are shown in Fig \ref{fig:residual_perfcompBoth}. First, we trained and tested our models with a dataset on where the introduced AWGN was varied from -60 dB to 10 dB. The classification performance is shown in Fig \ref{fig:residual_perfcompBoth} (left). The classifier without the skip connection could not learn from this noisy dataset, therefore, the performance remained poor throughout different SNR regions. However, the DRN model could learn features from the noisy dataset and the performance remained good throughout the SNR regions. 

We created another training dataset with a less noisy signal, where the AWGN was varied from -60 dB to -15 dB. With this dataset, the classifier without skip connection could learn necessary features to classify the signal. The performance is shown in Fig \ref{fig:residual_perfcompBoth} (right).

\subsection{Performance dependence on network depth}
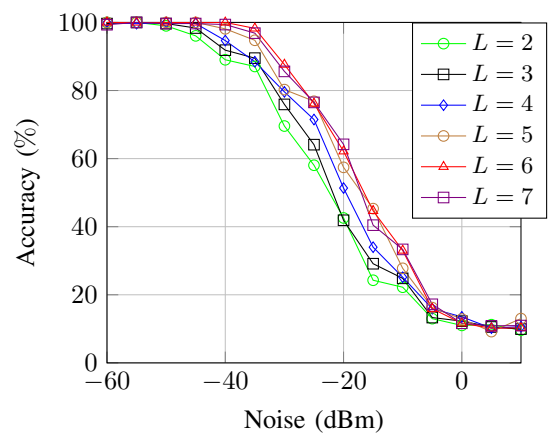
\begin{figure}[htbp]
\centering
\begin{minipage}[t][][b]{0.80\linewidth}
\centering
\begin{tikzpicture}
\begin{axis}
[legend style={at={(0.9,1)},anchor=north, font=\small},legend columns=1,
 width=\columnwidth,grid=both,ylabel=Accuracy ($\%$),xlabel=Noise (dBm),grid style={line width=.1pt, draw=gray!10},major grid style={line width=.2pt,draw=gray!50},xmin=-60,xmax=10,ymax=100, ymin=0,]

\addplot [mark=o, color=green] table [x=snr, y=s1, col sep=comma] {results/tikzplots/results_new/clc_layernumb.csv};\addlegendentry{$L=2$}

\addplot [mark=square, color=black] table [x=snr, y=s2, col sep=comma] {results/tikzplots/results_new/clc_layernumb.csv};\addlegendentry{$L=3$}

\addplot [mark=diamond, color=blue] table [x=snr, y=s3, col sep=comma] {results/tikzplots/results_new/clc_layernumb.csv};\addlegendentry{$L=4$}

\addplot [mark=o, color=brown] table [x=snr, y=s4, col sep=comma] {results/tikzplots/results_new/clc_layernumb.csv};\addlegendentry{$L=5$}

\addplot [mark=triangle, color=red] table [x=snr, y=s5, col sep=comma] {results/tikzplots/results_new/clc_layernumb.csv};\addlegendentry{$L=6$}

\addplot [mark=square, color=violet] table [x=snr, y=s6, col sep=comma] {results/tikzplots/results_new/clc_layernumb.csv};\addlegendentry{$L=7$}

\end{axis}
\end{tikzpicture} 
\end{minipage}
\caption{Classification performance vs network depth}
\label{fig:clc_netdepth}
\end{figure}

The classification performance generally improves with the increase in network depth. In this section, we show the dependency of the classification performance on the network depth. We varied the number of residual stacks (L) from 2 to 7 to visualize the impact of network depth on the classification performance. The average classification accuracy for different layers are plotted in Fig \ref{fig:clc_netdepth}. We can see that the accuracy increases with the increase in network depth. However, we haven't observed any improvement in classification performance after layer depth 6, the performance saturates after this layer depth.

\subsection{Performance comparison with other machine learning models}
\begin{figure}[t]
\centering
\begin{minipage}[b]{0.8\linewidth}
\centering
\begin{tikzpicture}
\begin{axis}
[legend style={at={(0.87,1.1)},anchor=north},legend columns=1,
 width=\columnwidth,grid=both,ylabel=F1 score ($\%$),xlabel=Noise (dBm),grid style={line width=.1pt, draw=gray!10},major grid style={line width=.2pt,draw=gray!50},xmin=-60,xmax=10,ymax=100, ymin=0,]

\addplot [mark=square, color=black] table [x=snr, y=f1rn, col sep=comma] {results/tikzplots/results_new/clc_comparison_F1.csv};\addlegendentry{$DRNN$}

\addplot [mark=diamond, color=red] table [x=snr, y=f1qt, col sep=comma] {results/tikzplots/results_new/clc_comparison_F1.csv};\addlegendentry{$FC-DNN$}

\addplot [mark=o, color=blue] table [x=snr, y=f1xg, col sep=comma] {results/tikzplots/results_new/clc_comparison_F1.csv};\addlegendentry{$XG$}

\addplot [mark=triangle, color=brown] table [x=snr, y=f1svm, col sep=comma] {results/tikzplots/results_new/clc_comparison_F1.csv};\addlegendentry{$SVM$}

\addplot [mark=oplus, color=violet] table [x=snr, y=f1rf, col sep=comma] {results/tikzplots/results_new/clc_comparison_F1.csv};\addlegendentry{$RF$}

\end{axis}
\end{tikzpicture} 
\end{minipage}
\caption{Performance comparison between different models under AWGN conditions}
\label{fig:comparison}
\end{figure}
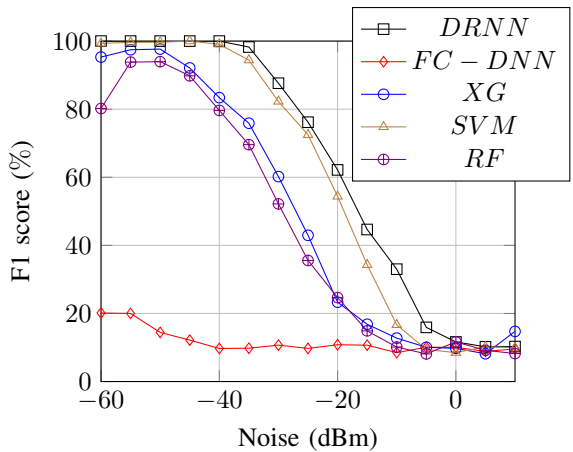

The classification performance of our framework is compared with other machine learning algorithms like \ac{svm}, Random Forest and \ac{xgboost} and the \ac{fc}-DNN model proposed in \cite{dronedatabase_qatarpaper}. The test is performed with the spectrogram dataset under AWGN conditions. For \ac{svm}, Random Forest and \ac{xgboost}, we have performed hyperparameter training using grid search. The classification performance for different frameworks is shown in Fig \ref{fig:comparison}. Our \ac{drnn} model provided the best performance compared to other models. At high SNRs, \ac{svm} with a linear kernel and penalty factor (C=1) provided the same F1 score as the \ac{drnn} model. At lower SNRs, our \ac{drnn} model performed better than the \ac{svm} framework. At AWGN -30 dBm (i.e. -9.4 dB \ac{snr}), our classifier provided nearly 5$\%$ better F1 score and at -15 dBm (i.e. -15 dB \ac{snr}), it gave nearly 10$\%$ better F1 score compared to the SVM model.    
XGBoost with a max depth of 10 and random forest with 200 decision trees provided a lower F1-score compared the \ac{svm} and \ac{drnn} model. On the contrary, \ac{fc}-DNN model showed a poor performance compared to all other frameworks. This clearly shows that a simple 3 layer \ac{fc}-DNN model is not appropriate for classifying noisy spectrogram dataset.

\subsection{Simultaneous multidrone classification}
\begin{figure}[htbp]
\centering
\begin{minipage}[b]{0.39\linewidth}
\includegraphics[trim=0.1cm 0.1cm 0.1cm 0.01cm,clip=true,width=\columnwidth]{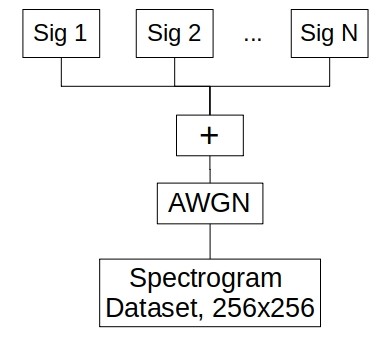}
\subcaption{Simulation schematic}
\label{fig:sigrecsim}
\end{minipage}
\hfill
\begin{minipage}[b]{0.45\linewidth}
\includegraphics[trim=0.2cm 0.2cm 0.2cm 0.2cm,clip=true,width=\columnwidth]{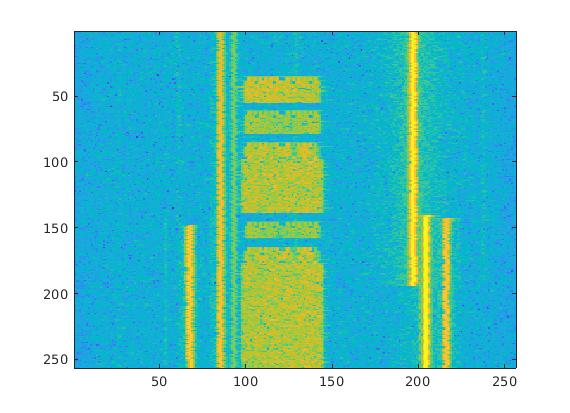}
\subcaption{Simultaneous presence of 7 signals}
\label{fig:multisig_addition}
\end{minipage}
\caption{Simultaneous multiple signal classification experiment}
\label{fig:multisigschema}
\end{figure}

In this section, we investigate the problem of simultaneous multiple drone classification. The simulation schematic of the dataset preparation is shown in Fig \ref{fig:sigrecsim}. We have added the signals on Matlab to have all different possible combinations and fed through an AWGN channel. One example is shown in Fig \ref{fig:multisig_addition}, where seven signals are present in the spectrogram dataset. 

\begin{figure}[htbp]
\centering
\begin{minipage}[b]{0.70\linewidth}
\centering
\begin{tikzpicture}
\begin{axis}
[legend style={at={(0.3,0.6)},anchor=north, font=\small},
 legend columns=1,
 width=\columnwidth,
 grid=both,
 ylabel=F1 Score ($\%$),
 xlabel=Noise (dBm),
 grid style={line width=.1pt, draw=gray!10},
 major grid style={line width=.2pt,draw=gray!50},
 xmin=-50,xmax=-10,ymax=100, ymin=0,]

\addplot [mark=square, color=black] table [x=snr, y=F1_7, col sep=comma] {results/tikzplots/results_new/clc_multisig_F1.csv};\addlegendentry{$Nb src=7$}

\addplot [mark=diamond, color=red] table [x=snr, y=F1_5, col sep=comma] {results/tikzplots/results_new/clc_multisig_F1.csv};\addlegendentry{$Nb src=5$}

\addplot [mark=o, color=blue] table [x=snr, y=F1_3, col sep=comma] {results/tikzplots/results_new/clc_multisig_F1.csv};\addlegendentry{$Nb src=3$}

\addplot [mark=triangle, color=brown] table [x=snr, y=F1_2, col sep=comma] {results/tikzplots/results_new/clc_multisig_F1.csv};\addlegendentry{$Nb src=2$}

\end{axis}
\end{tikzpicture}
\end{minipage}
\caption{Simultanous multi-drone classification performance under AWGN conditions}
\label{fig:multidroneclassification}
\end{figure}
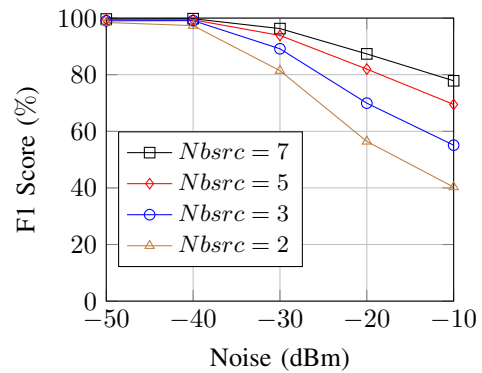

The classification performance for simultaneous multi-drone is shown in Fig \ref{fig:multidroneclassification}. We have used 0.5 as the decision threshold. We can see that the classification performance is almost the same as with the single drone classification. We obtained F1 score ranging from 97.3 to 99.7 $\%$ for the introduced AWGN ranging from -40 to -50 dBm. We did not observe any significant decrease in the classification performance with the increase in the number of the source. On the contrary, an increase in classification performance can be observed with the increase in source numbers. Since within the dataset we mostly have narrowband signals and they are less impacted with high noise compared to the wideband signals, the combination of the signal showed better performance on average compared to the single drone scenario. The classification performance of each class remained almost the same for the simultaneous multi-class scenario as the singular counterpart. The result shows that our classifier does not only classify multiple drones simultaneously at lower SNR regions perfectly, it can also distinguish and classify the drone signals in the presence of WiFi communication.

\section*{Conclusion}
In this paper, we have proposed a deep residual network to classify several drone signals in a single drone and simultaneous multi-drone scenario. We have created a dataset using nine commercial drones and WiFi signals and evaluated the classification performance in AWGN and multipath environments. The importance of residual mapping and the dependence of the network depth on the classification performance is presented. Along with that, the class activation map is visualized. Furthermore, we have compared the classification performance of our model with other existing frameworks. Our proposed method outperformed the existing RF-based signal classification models and other standard wireless identification techniques by a good margin. At around 0 dB SNR, we achieved nearly 99$\%$ classification accuracy for both single and simultaneous multi-drone scenario. For the single drone scenario, our classifier provided around 5$\%$ better F1 score compared to the existing framework at -10 dB \ac{snr}. In the future  work, we are going to investigate the classification performance in a frequency overlap scenario. Along with that, we are going to investigate combined  RF  signal  detection,  localization  and  classification using  the  latest  computer  vision  methods  and  compare the performance  with  the  two-stage  detection  and  classification method.

\bibliographystyle{IEEEtran}
\bibliography{drone_main}

\vspace{4pt}

\end{document}